\documentclass{moriond}
\usepackage{amsmath}
\usepackage{longtable} 
\usepackage[table]{xcolor}
\usepackage{hyperref}

\def\be{\begin{equation}}
\def\ee{\end{equation}}
\def\bea{\begin{eqnarray}}
\def\eea{\end{eqnarray}}

\usepackage{color}
\definecolor{mygreen}{RGB}{143, 176, 50}
\definecolor{myorange}{RGB}{225, 156, 36}
\definecolor{myblue}{RGB}{94, 129, 181}

\newcommand{\Photo}{}

\begin{document}
\vspace*{4cm}
\title{Direct comparison of  global fits to the $B \to K^* \mu^+ \mu^-$  data assuming  hadronic corrections or new physics}

\author{S.~Neshatpour$^{1,\;}$\footnote{Speaker}, V.G.~Chobanova$^2$, T.~Hurth$^3$, F.~Mahmoudi$^4$, and D.~Mart\'inez Santos$^2$}
\address{$^1$ School of Particles and Accelerators, Institute for Research in Fundamental Sciences (IPM)\\ 
P.O. Box 19395-5531, Tehran, Iran\\
$^2$ Instituto Galego de F\'isica de Altas Enerx\'ias, Universidade de Santiago de Compotela, Spain\\
$^3$ PRISMA Cluster of Excellence and  Institute for Physics (THEP) Johannes Gutenberg University, D-55099 Mainz, Germany\\
$^4$ Univ Lyon, Univ Lyon 1, ENS de Lyon, CNRS, Centre de Recherche Astrophysique de Lyon UMR5574, F-69230 Saint-Genis-Laval, France\\
Theoretical Physics Department, CERN, CH-1211 Geneva 23, Switzerland
}

%
%
%

\maketitle\abstracts{
The LHCb measurements on the $B \to K^* \mu^+ \mu^-$ angular observables 
have shown slight deviations from their Standard Model
predictions. 
The significance of the deviations in the $B \to K^* \mu^+ \mu^-$ decay
depends on the assumptions on the size 
of the non-factorisable power corrections. 
At present, there are no theoretical predictions on the size of these power corrections
in order to identify whether the reason behind these anomalies is due to unknown hadronic 
corrections or New Physics effects. We have performed a statistical comparison 
of fitting the data to each of the possible explanations.
}

\section{Introduction}
The LHCb measurements on the angular observables of the $B\to K^* \mu^+ \mu^-$ decay was presented in 2013 with 1 fb$^{-1}$ of data\!~\cite{Aaij:2013qta}  
and were mostly consistent with the SM predictions, with slight deviations in specific bins/observables. 
The largest tension was observed in the $P_5^\prime$ angular
observable with $3.7\sigma$ significance for the bin with $q^2 \in [4.30,8.68]$ GeV$^2$. 
Less significant tensions were observed in some of the other angular observables such as $P_2$.
The $P_5^\prime$ tension was later reconfirmed by LHCb with 3 fb$^{-1}$ of data\!~\cite{Aaij:2015oid}, in the finer  $[4.0,6.0]$ and $[6.0,8.0]$ GeV$^2$ bins, with $2.8$ and $3.0\sigma$ significance, respectively.
More recently, the Belle collaboration\!~\cite{Abdesselam:2016llu} as well as the ATLAS collaboration\!~\cite{ATLAS-CONF-2017-023} have reaffirmed this tension, although with 
less significance which is mostly due to the larger experimental uncertainties compared to LHCb.
However, the experimental result from the CMS collaboration\!~\cite{CMS-PAS-BPH-15-008} does not confirm the tension in $P_5^\prime$.

With several experimental measurements on $P_5^\prime$ which mostly indicate deviations from the SM prediction, 
it seems unlikely for statistical fluctuations to be the source of the tension and 
the most likely explanation would either be underestimated theoretical (hadronic) uncertainties or New Physics (NP) effects.

The LHCb measurements with 3 fb$^{-1}$ dataset for other $b\to s \bar \ell \ell$ transitions indicate further deviations with the SM predictions at 2-4$\sigma$
significance level in observables such as BR$(B_s \to \phi \mu^+ \mu^-)$~\cite{Aaij:2015esa} and also
$R_K\equiv {\rm BR}(B\to K^+ \mu^+ \mu^-)/ {\rm BR}(B \to K^+ e^+ e^-)$~\cite{Aaij:2014ora}$^{,}$~\footnote{Recently, 
a similar measurement on $R_{K^*}$ was presented by the LHCb collaboration\!~\cite{Aaij:2017vbb}.}. 
Interestingly, the latter tensions, as well as the anomalies in the angular observables can all be explained with a common NP effect, namely about 25\% reduction in the $C_9^{(\mu)}$
Wilson coefficient\!~\cite{Hurth:2014vma,Altmannshofer:2014rta,Descotes-Genon:2015uva}.

Besides the $R_K$ observable which is very precisely predicted in the SM, the other observables suffer from hadronic effects.
The standard method for calculating the hadronic effects in the low $q^2$ region for the exclusive $B\to K^* \bar \ell \ell$ 
decay is the QCD factorisation framework where an expansion of 
$\Lambda/m_b$ is employed. However, higher powers of $\Lambda/m_b$ remain unknown and so far are only ``guesstimated''.
The significance of the anomalies to a large extent depends on the precise treatment of these non-factorisable power corrections\!~\cite{Hurth:2016fbr,Mahmoudi:2016mgr}.
In the absence of concrete estimations of the power corrections, we make  
a statistical comparison between a NP fit and a hadronic power corrections fit 
to the $B \to K^* \mu^+ \mu^-$ measurements\!~\cite{Chobanova:2017ghn}.

\section{Nonfactorisable power corrections vs NP}\label{sec:theory}
The $b \to s \bar \ell \ell$ transitions can be described via an effective Hamiltonian which can be formally separated into 
a hadronic and a semileptonic part\!~\cite{Jager:2012uw}:
\begin{equation}
 {\cal H}_{\rm eff} = {\cal H}_{\rm eff}^{\rm had} + {\cal H}_{\rm eff}^{\rm sl}\,,
\end{equation}
where
\begin{equation}
 {\cal H}_{\rm eff}^{\rm had}=-\frac{4G_F}{\sqrt{2}}V_{tb}V_{ts}^*\sum_{i=1,\ldots,6,8}C_i\; O_i\,, 
 \quad {\cal H}_{\rm eff}^{\rm sl}=-\frac{4G_F}{\sqrt{2}}V_{tb}V_{ts}^*
 \sum_{i=7,9,10,S,P,T}(C_i\; O_i + C^\prime_i\; O^\prime_i)\,.
\end{equation}
For the exclusive decays $B\to K^* \mu^+ \mu^-$ and $B_s \to \phi \mu^+ \mu^-$, the semileptonic part of the
Hamiltonian,  ${\cal H}_{\rm eff}^{\rm sl}$, which accounts for the dominant contribution, can be described by seven independent (helicity) form factors 
$\tilde{S}, \tilde{V}_\lambda, \tilde{T}_\lambda$, with $\lambda=\pm1,0$ indicating the helicities. 
The exclusive $B\to V \bar \ell \ell$ decay, where $V$ is a vector meson can be described (in the SM) by seven helicity amplitudes:
\begin{equation}\label{eq:HV}    
\begin{split}
  H_V(\lambda) &=-i\, N^\prime \Big\{ C_{9}^{\rm eff} \tilde{V}_{\lambda}(q^2) 
      + \frac{m_B^2}{q^2} \Big[\frac{2\,\hat m_b}{m_B} C_{7}^{\rm eff} \tilde{T}_{\lambda}(q^2)
      - 16 \pi^2 {\cal N}_\lambda(q^2) \Big] \Big\}\, , \\
  H_A(\lambda) &= -i\, N^\prime C_{10}  \tilde{V}_{\lambda} \, , \\
  H_P &= i\, N^\prime \Big\{  \frac{2\,m_\ell \hat m_b}{q^2} 
                     C_{10}( 1 + \frac{m_s}{m_b} )\tilde{S}\Big\}\,,
\end{split}
\end{equation}
where the effective part of $C_9^{\rm eff}\left(\equiv C_9+Y(q^2)\right)$ as well as the 
nonfactorisable contribution ${\cal N}_\lambda(q^2) \left(\equiv \text{LO in QCDf} + h_\lambda (q^2) \right)$ arise 
from the hadronic part of the Hamiltonian through the emission of a photon which itself turns into a lepton pair.
Due to the vectorial coupling of the photon to the lepton pair, the contributions of ${\cal H}_{\rm eff}^{\rm had}$ appear in
the vectorial helicity amplitude $H_V(\lambda)$.
It is due to the similar effect from the short-distance $C_9$ (and $C_7$) of ${\cal H}_{\rm eff}^{\rm sl}$ and the long-distance contribution from ${\cal H}_{\rm eff}^{\rm had}$ that
there is an ambiguity in separating NP effects of the type $C_9^{\rm NP}$ (and $C_7^{\rm NP}$) from nonfactorisable hadronic contributions.

The nonfactorisable contribution ${\cal N}_\lambda(q^2)$ is known at leading order in $\Lambda/m_b$ from QCDf calculations 
while higher powers $h_\lambda (q^2)$ are only partially known\!~\cite{Khodjamirian:2012rm} and can only be guesstimated.
These power corrections are usually assumed to be 10\%, 20\%, etc. of the leading order nonfactorisable contribution.
However, instead of making an ansatz on the size of the power corrections they can be fitted to the experimental data\!~\cite{Ciuchini:2015qxb}.
One rather general description of the power corrections is through a  $q^2$-dependent expression
\begin{equation}\label{eq:hlambda}
  h_\lambda(q^2)= h_\lambda^{(0)} + \frac{q^2}{1 {\rm GeV}^2}h_\lambda^{(1)} 
  + \frac{q^4}{1 {\rm GeV}^4}h_\lambda^{(2)}\,,
\end{equation}
where considering  each $h_\lambda^{(0,1,2)}$ to be a complex number, is described by 18 free (unknown) real parameters.
It might seem that the hadronic power corrections and  NP effects
can be differentiated by the lack of $q^2$-dependence in the $C_9^{\rm NP}$ (and $C_7^{\rm NP}$) expressions.
However, this is not true since the latter are multiplied by $q^2$-dependent form factors $\tilde{V}(q^2)$ ($\tilde{T}(q^2)$)
and it has been shown\!~\cite{Chobanova:2017ghn} that the NP effect in $C_9$ (and $C_7$) can be embedded in the more general case of hadronic power corrections
(Eq.~\ref{eq:hlambda}) and hence it is reasonable to make a statistical comparison of a hadronic fit and a NP fit of $C_9$ (and $C_7$) to the $B\to K^* \mu^+ \mu^-$ data.

\section{Results}
\begin{figure}[t!]
\centering
\includegraphics[width=0.55\textwidth]{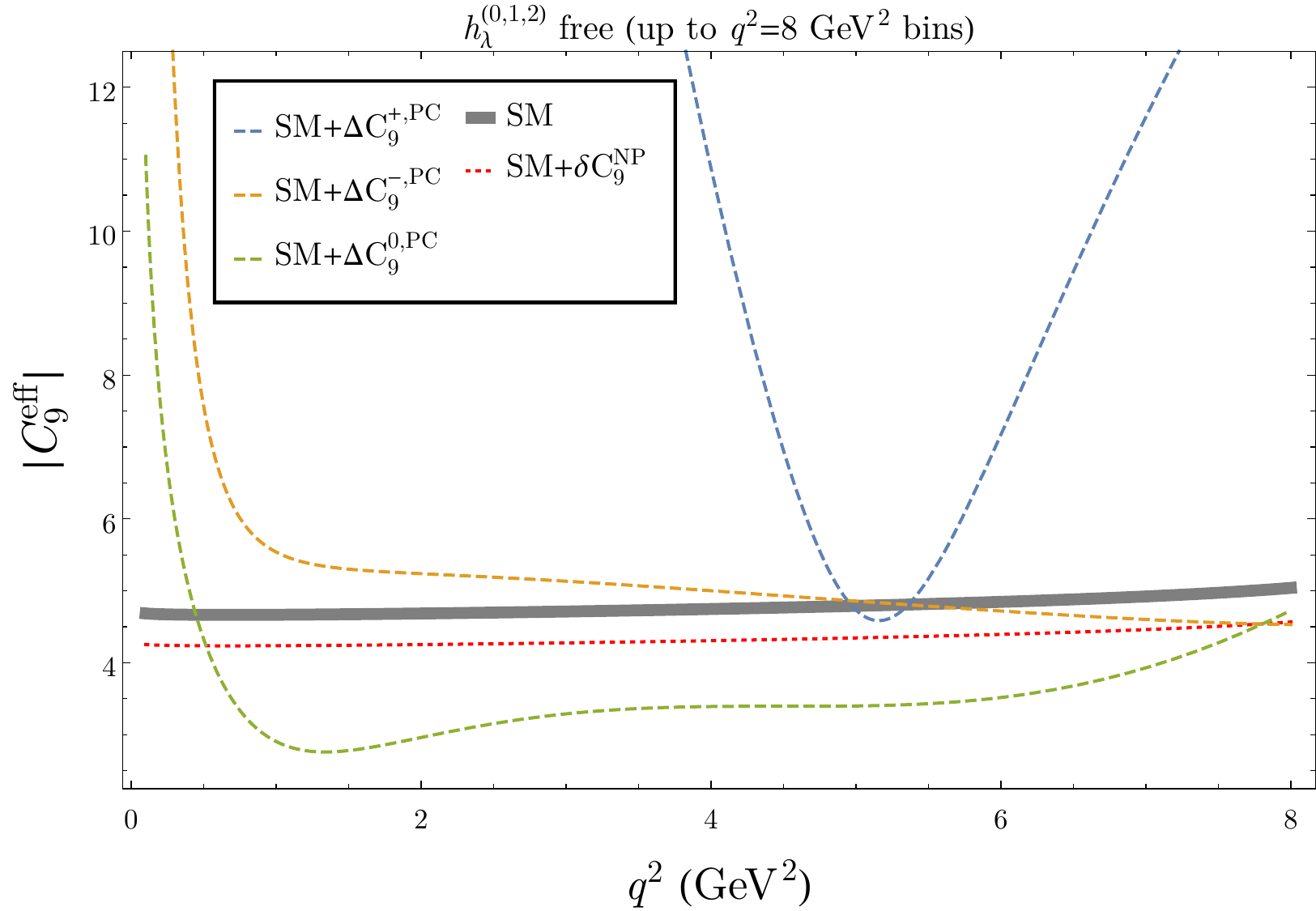}
\caption{The solid line corresponds to the SM value of $C_9^{\rm eff}(q^2)$.
The dotted line corresponds to {\color{red}$C_9^{\rm eff, SM}(q^2)+\delta C_9^{\rm NP}$}, 
where  $\delta C_9^{\rm NP}=-0.97-i\times2.13\;$ is the best fit 
value of the NP fit for $C_9$.
The hadronic power corrections fit results are demonstrated by dashed lines via
\textcolor{myblue}{$C_9^{\rm eff, SM}+\Delta C_9^{+,PC}$}, 
\textcolor{myorange}{$C_9^{\rm eff, SM}+\Delta C_9^{-,PC}$} 
and \textcolor{mygreen}{$C_9^{\rm eff, SM}+\Delta C_9^{0,PC}$}.
\label{fig:DeltaC9}}
\end{figure}
The fits are done by the MINUIT minimisation tool with theoretical predictions from SuperIso v3.6\!~\cite{Mahmoudi:2007vz,Mahmoudi:2008tp}
and considering CP-averaged $B\to K^* \mu^+ \mu^-$ observables at $q^2<8$ GeV$^2$.
For the NP scenarios, we have fitted  $C_9$ (and $C_7$) which assuming complex Wilson coefficient(s)
involves 2~(4) free parameters. For the hadronic power corrections, we have fitted the 18 free parameters, assuming complex values for $h_{\pm,0}^{(0,1,2)}$.
To compare the NP fit and the hadronic corrections fit, we demonstrate (Fig.~\ref{fig:DeltaC9}) the latter through $q^2$-dependent shifts to $C_9^{\rm eff}$ via
\begin{align}
 \Delta C_9^{\lambda,\rm{PC}}= -16 \pi^2 \frac{m_B^2}{q^2} \frac{h_\lambda (q^2)}{\tilde{V}_\lambda(q^2)}.
\end{align}
In principle, the hadronic power corrections can mimic the NP scenario. However, this would be rather unlikely since for that, 
the three distinct hadronic corrections should conspire to imitate the $C_9$ effect 
(i.e. in Fig.~\ref{fig:DeltaC9}, the three dashed lines indicating the power corrections 
would need to coincide with the dotted line corresponding to  $C_9^{\rm NP}$).

The various scenarios can be compared through likelihood ratio tests via Wilks' theorem.
Considering the difference in number of parameters between two scenarios and taking $\Delta \chi^2$, the $p$-values are obtained.
The $p$-values imply the significance of adding parameters to go from one nested scenario to a more general case.
\begin{table}[bh!]
\begin{center}
\setlength\extrarowheight{1pt}
\scalebox{0.8}{
\begin{tabular}{|c|c|c|c|c|c|} \hline
& $\delta C_9$ & $\delta C_7,\delta C_9$ & Hadronic\\ \hline
plain SM                & $3.7\times 10^{-5}(4.1\sigma)$ & $6.3\times10^{-5}(4.0\sigma)$ & $6.1\times 10^{-3}(2.7\sigma)$\\
$\delta C_9$            & --                             & $0.13 (1.5\sigma)$            & $0.45(0.76\sigma)$\\
$\delta C_7,\delta C_9$ & --                             & --                            & $0.61(0.52\sigma)$ \\
\hline
\end{tabular} }
\caption{$p$-values and significances of adding parameters to go from one scenario to another using Wilks' theorem.
\label{tab:Wilks8}}
\end{center} 
\end{table}  
From Table~\ref{tab:Wilks8}, it can be seen that adding the hadronic parameters (16 more parameters) compared to the $C_9^{\rm NP}$ scenario does not really improve the fits
(the fit is only improved by $0.76\sigma$ significance) and the NP  explanation remains as a justified option for interpreting the tensions in the angular observables.

\section{Conclusions and outlook}
Due to the embedding of NP effects in the more general case of hadronic power corrections, 
any NP effect could be mimicked by some hadronic effect. Hence in principle, it is not possible to rule out hadronic contributions 
in favour of NP effects as long as ``embedding breaking'' observables such as  CP violating  $B\to K^* \mu^+ \mu^-$ observables or 
lepton flavour universality breaking observables such as $R_K$ are not considered.
However, we have explicitly shown that the additional hadronic parameters do not improve the fit.
This is due to the fact that  with the current experimental data, the fit results of the power corrections 
are mostly consistent with zero in the $1\sigma$ range~\cite{Chobanova:2017ghn}.
Thus, our fits show that the NP interpretation of the anomalies in the $B\to K^* \mu^+ \mu^-$ angular observables
is still a tenable option.

While with the current data, the hadronic corrections fit 
has a rather mild $q^2$-dependence, this should not be misinterpreted as a proof
for the NP option because this behaviour might be due to a possible smearing 
of resonances and could be changed with a smaller binning of the experimental data.

The LHCb upgrade is expected to yield 300 fb$^{-1}$ of data at the high-luminosity LHC. 
Assuming the current central values of $B\to K^* \mu^+ \mu^-$ measurements remain and that the experimental errors get reduced by a factor of 10,
a similar statistical comparison 
would strongly  rule out the NP scenario with 34$\sigma$ significance.

\section*{Acknowledgments}
The speaker, SN, is grateful to the organisers of the fruitful Rencontres de Moriond
QCD 2017  conference, and appreciates very much the financial support and opportunity to present a talk.

\end{document}